\pdfoutput=1
\newtheorem{theorem}{Theorem}

\newtheorem{definition}{Definition} 

\documentclass{jpconf}

\usepackage{amsfonts}

\usepackage[latin1]{inputenc}
\usepackage{graphicx}

\begin{document}

\title{
Local gauge theory and coarse graining
}
\author{
José A. Zapata
}
\address{Centro de Ciencias Matemáticas, UNAM,
		58190 Morelia, Mexico}
\ead{zapata@matmor.unam.mx}
\begin{abstract}
Within the discrete gauge theory 
which is the basis of spin foam models, the problem of macroscopically faithful coarse graining is studied. 
Macroscopic data is identified; it contains the holonomy evaluation along a discrete set of loops and the homotopy classes of certain maps. When two configurations share this data they are related by a local deformation. The interpretation is that such configurations differ by  ``microscopic details''. 
In many cases the homotopy type of the relevant maps is trivial for every connection; two important cases in which the homotopy data is composed by a set of integer numbers  are: (i) a two dimensional base manifold and structure group $U(1)$, (ii) a four dimensional base manifold and structure group $SU(2)$. These cases are relevant for spin foam models of two dimensional gravity and four dimensional gravity respectively.
This result suggests that if spin foam models for 
two-dimensional and four-dimensional gravity 
are modified to include all the relevant macroscopic degrees of freedom 
--the complete collection of macroscopic variables necessary to ensure faithful coarse graining--, 
then they could provide appropriate effective theories at a given scale. 
\end{abstract}

\section{Connections (modulo gauge) in terms of local holonomy variables}
Let $(E, \pi , M)$ be a principal $G$-bundle over a $d$-dimensional base space $M$ 
and ${\cal A}_\pi$ be the space of connections on it. 

In order to introduce a concept of measuring scale and locality we use a smooth triangulation $\Delta$ of $M$; alternatively, we can use a cellular decomposition that refines a smooth triangulation. 
This decomposition is inherited by the bundle (Figure 1.a). 
\begin{figure}[h!]
  \centering
{%
      \includegraphics[width=0.7\textwidth]{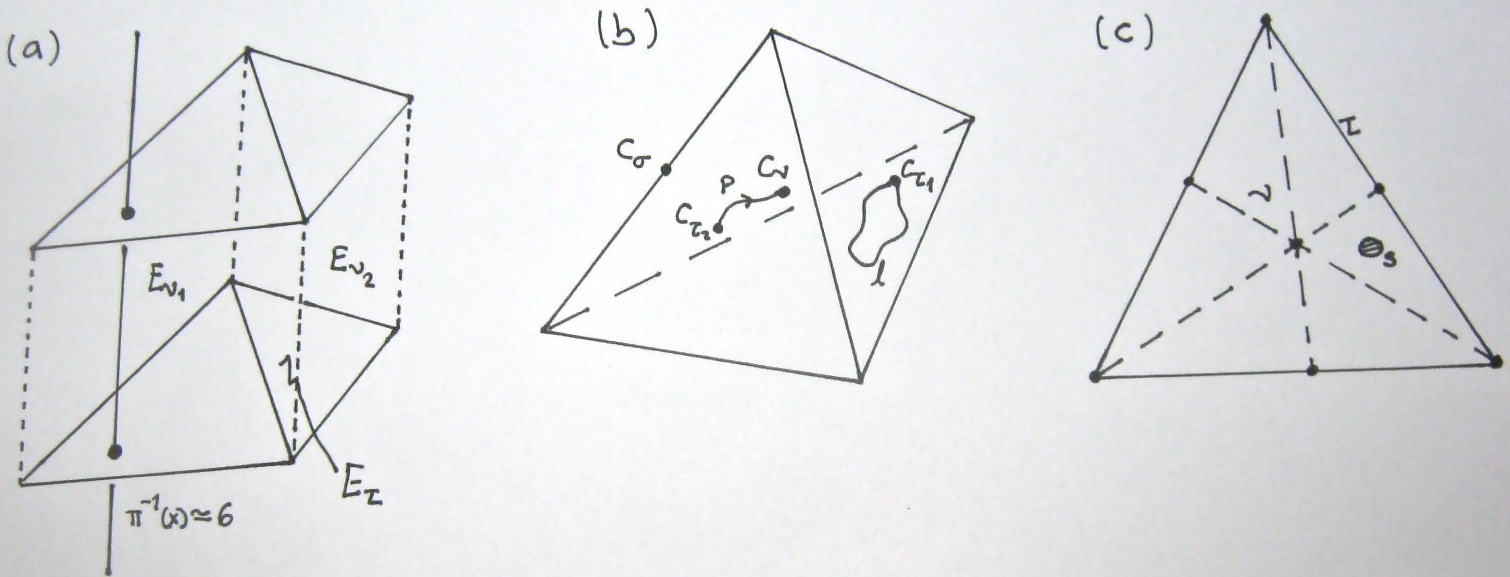}}
\caption{
(a) The smooth $G$-bundles $\pi_{\nu_1}$ and $\pi_{\nu_2}$ intersect at the smooth $G$-bundle $\pi_{\tau}$. \\
(b)  Paths and loops in ${\cal P}(\nu)$. \\
(c) ${\rm Sd} (\nu)$, the baricentric subdivision of $\nu$. We also show $s$, the support of $B$ in the example. 
}
\end{figure}
\begin{definition}
\label{DeltaSmooth}
Decomposition of the bundle (see figure 1.a):
\begin{itemize}
\item 
The  principal  $G$-bundle over $M$, $(E_\Delta, \pi_\Delta , M_\Delta)$, is called $\Delta$-smooth if 
its restriction to $E_\nu= \pi^{-1}_\Delta(\nu)$ for each simplex $\nu$ of the triangulation is the restriction of a smooth bundle to an open neighborhood of 
$E_\nu\subset E_\Delta$. 
\item 
${\cal A}_{\pi_\tau}^\infty$ is the space of smooth connections 
in the bundle $\pi_\tau = \pi |_{\tau \in (|\Delta | , \phi)}$ (restrictions to $\pi_\tau$ 
of connections which are smooth in an open neighborhood of the subbundle $\pi_\tau$ of $\pi$, where $\tau$ is a simplex in the triangulation). 
\item 
$\bar{\times}_{\tau \in (|\Delta | , \phi)}  {\cal A}_{\pi_\tau}^\infty$ 
is the subset of the cartesian product 
defined by the compatibility condition 
$\sigma \subset \tau \Rightarrow A_\sigma = A_\tau |_\sigma$ for every two simplices of the triangulation. 
We will write \\
${\cal A}_\pi^{\Delta\mbox{-}\infty} = 
\bar{\times}_{\tau \in (|\Delta | , \phi)}  {\cal A}_{\pi_\tau}^\infty$. 
\item 
A $\Delta$-smooth bundle map is a 
homeomorphism of the total space $\tilde{f}: E \to E$ 
which preserves each subbundle $\pi_\tau$ and 
is the restriction to $\pi_\tau$ of a smooth bundle map on an open neighborhood of it. 
Clearly a $\Delta$-smooth bundle map acts on connections by pull back and sends 
${\cal A}_\pi^{\Delta\mbox{-}\infty}$ to itself. 
\item 
${\cal G}_{\star, \pi}^{\Delta\mbox{-}\infty}$ 
is the group of $\Delta$-smooth bundle equivalence maps which consists of 
$\Delta$-smooth bundle maps which induce the identity map on the base space $M$. 
We will write 
${\cal A/G}_{\star , \pi}^{\Delta\mbox{-}\infty} = 
{\cal A}_\pi^{\Delta\mbox{-}\infty} / {\cal G}_{\star, \pi}^{\Delta\mbox{-}\infty}$. 

\end{itemize}
\end{definition}

Now we present local holonomy data for the bundle and the connection. It is the application of a construction of Barrett and Kobayashi to the subbundles over the simplices. 

In simplex $\nu$ of the triangulation choose its baricenter $C\nu$ as base point, 
and consider the space ${\cal L}(C\nu, \nu)$ 
of piecewise smooth oriented $C\nu$-based loops modulo reparametrization and modulo retracing. It turns out that ${\cal L}(C\nu, \nu)$ is a group. 
Once we have identified $G$ with $\pi^{-1}(C\nu)$
the holonomy of a given connection $A \in {\cal A}_{\pi_\nu}^\infty$
around a loop $l \in {\cal L}(C\nu, \nu)$ is the group element 
${\rm Hol}(l , A) \in G$. Moreover, if we fix the connection the holonomy map gives us a group homomorphism 
\[
{\rm Hol}_A \equiv {\rm Hol}( \cdot , A): {\cal L}(C\nu, \nu) \to G . 
\]
Gauge transformations act on holonomies 
only at the base point and the action is by conjugation. 

Holonomy evaluations 
are specified by points of the space
${\cal A/G}_{\star , \pi_\nu}$, the space of connections modulo the group of gauge transformations whose restriction to the fiber $\pi^{-1}(C\nu)$ is the identity. 
Holonomy evaluations 
can be used to construct a trivialization of the bundle over $\nu$, and they 
characterize the connection (modulo gauge); this is an application of 
a theorem of Barrett and Kobayashi \cite{Barrett:1991aj} to the simple case where the base space is a topological disc. 

The theorem provides a construction of 
the bundle and the connection. Here is a sketch of the main idea. The total space is given by 
\[
E_\nu = G \times {\cal P}_{C_\nu} / \sim_{{\rm Hol}^\nu}
\]
where ${\cal P}_{C_\nu}$ is the space of paths in $\nu$ (modulo reparametrization and retracing) whose source is $C_\nu$. The idea behind the equivalence relation is that a pair $(g, p) \in G \times {\cal P}_{C_\nu}$ gives a point on the fiber over $t(p)$, the target of $p$; 
$g$ specifies a point on the fiber at the beginning of the path and it is parallel transported to the end of the path. One only has to acknowledge that 
$(g_1, p_1)$ and $(g_2, p_2)$ determine the same point if $t(p_1) = t(p_2)$ and 
the parallel transports agree, 
${\rm PT}_{p_1}g_1 = {\rm PT}_{p_2}g_2$. This last condition can also be written as 
$g_1{\rm H}^\nu(p_2^{-1}\circ p_1) = g_2$. 

If this construction is applied to each simplex of the smooth triangulation, the connection and the bundle will be described by means of holonomies. However, a collection of such holonomy evaluations for each simplex will be related to a bundle and a connection on it only if certain compatibility conditions hold. The pasting of descriptions over the different simplices needs extra data. The extra data glues the bundle over a simplex to the bundle over any simplex that intersect it. 

Consider two simplices of the smooth triangulation such that $\tau \subset \nu$. Thus, for each point $x \in \tau$ the above construction gives two constructions of $\pi_x$: 
$E_\tau|_x$ and $E_\nu|_x$. The glue for these bundles is again parallel transport. There are two groups ${\cal L}(C\tau, \tau)$ and ${\cal L}(C\nu, \nu)$; and there are also paths in $\nu$ between $C\tau$ and $C\nu$.  
The three spaces can be considered together forming a semigroup; two of its elements can be composed only if the target of the first is the same as the source of the second. 
Consider all the subsimplices of $\nu$, $\tau_i \subset \nu$. 
The semigroup of paths in $\nu$ which have sources and targets at $C_\nu$ or any of the $C_{\tau_i}$ will be called ${\cal P}(\nu)$ (see figure 1.b). 
Notice that this semigroup contains ${\cal L}(C\nu, \nu)$ and 
${\cal L}(C\tau_i, \tau_i)$ for each $\tau_i \subset \nu$. 
Once the fibers over $C_\nu$ and all the $C_{\tau_i}$s have been identified with $G$, parallel transport is described by 
the semigroup homomorphism 
\[
{\rm PT}^\nu: {\cal P}(\nu) \to G . 
\] 
${\rm PT}^\nu$ 
contains all the information in ${\rm H}^\nu$ and each of the 
${\rm H}^{\tau_i}$; in addition, it 
induces a further equivalence relation that glues the bundle over $\nu$ with the bundles over each $\tau_i$. 
For example, $(g_1, p_1) \in E_\nu|_x$ is equivalent to $(g_2, p_2) \in E_\tau|_x$ if 
$g_1{\rm PT}^\nu(p_2^{-1} \circ p_1)= g_2$. 

We can extend this construction to glue the bundles over all the simplices of the smooth triangulation to obtain a variation of the 
Barrett and Kobayashi theorem which is tailored to the sense of locality given by the triangulation. 
The ingredients are ${\cal P}(\Delta)$ --the semigroup associated to the smooth triangulation of the manifold- and a parallel transport map which will be called simply 
${\rm PT}$. 
${\cal P}(\Delta)$ is 
generated by composing paths (modulo reparamerization and retracing) 
contained in some simplex of the triangulation and with source and target being baricenters of simplices of the triangulation. 
${\rm PT}$ is required to be a 
$\Delta$-smooth 
semigroup homomorphism. The smoothness condition means that its restriction to paths contained in 
any simplex is the restriction of a map on paths contained in an open neighborhood of the simplex which is smooth in the sense of Barrett \cite{Barrett:1991aj}. 
A detailed study of the abelian case was presented at \cite{DZ}. 
\begin{theorem}
\label{ReconstructionSmoothTriang}
Let $M$ be a manifold, $\Delta$ a smooth triangulation of it, and let 
${\rm PT}: {\cal P}(\Delta) \to G$ 
be a $\Delta$-smooth semigroup homomorphism. Then, there is a 
$\Delta$-smooth 
$G$-bundle $(E, \pi_\Delta, M)$, 
a collection of points $\{b_\nu \in \pi^{-1} (C\nu)\}_{\nu \in \Delta}$, and a connection 
$A \in {\cal A}_{\pi_\Delta}$ such that ${\rm PT} = {\rm PT}_A $. 
The bundle and the connection are unique up to a bundle equivalence transformation. 
\end{theorem}
\section{Macroscopic variables and faithful coarse graining}

The last theorem describes the bundle and the connection in terms of a parallel transport map. 

Recall that since bundles over discs are trivial and simplices have the topology of the disc. 
Different parallel transport maps may yield inequivalent bundles only because they glue {\em standard building blocks} in inequivalent ways determined by the parallel transport data. 

Each ${\cal P}(\nu)$ is decomposed into subsemigroups that intersect each other as dictated by the simplicial structure of triangulation
\[
\nu_1 \cap \nu_2 = \tau \Rightarrow {\cal P}(\nu_1) \cap {\cal P}(\nu_2) = {\cal P}(\tau). 
\]
This property lets us treat one simplex at a time, and see non trivial gluing 
as caused by the gluing of individual bundles over simplices with the bundles over their boundary faces. 

If $\tau \subset \partial \nu$ then Theorem \ref{ReconstructionSmoothTriang} provides 
$\pi_\nu$, $\pi_\tau$ and a gluing map $I_{\tau \nu}: \pi_\tau \to \pi_\nu$ determined by ${\rm PT}^\nu$. 
We will show that there is macroscopic data that can be extracted from ${\rm PT}^\nu$ assuring that if 
$\hbox{Data}_\Delta({\rm PT}_1^\nu) = \hbox{Data}_\Delta({\rm PT}_2^\nu)$ then 
$I_{\tau \nu}$ is equivalent to 
$I_{\tau \nu}$. 
A realization of this goal in the abelian case was presented at \cite{DZ}. 

Below is an example of inequivalent gluings.
It shows that $\hbox{Data}_\Delta({\rm PT}^\nu)$ must include some extra data apart from a set of parallel transport evaluations along a discrete collection of paths and loops. 

{\bf Example:}\\
\noindent
{\small \em 
Let $G = U(1)$ and $M \sim S^2$ with $\Delta$ a smooth triangulation of it; also let $\nu \in \Delta$ be a triangle and $\tau \in \Delta$ be one of its one dimensional faces. 
The baricentric subdivision of $\nu$, ${\rm Sd} (\nu)$, is shown in figure 1.c. 

In $S^2$ we have a parallel transport map ${\rm PT}_1$ such that 
${\rm PT}_1^\nu$ is a flat parallel transport (all holonomies are the identity). In an appropriate local trivialization over $\nu$, ${\rm PT}_1^\nu$ would be described by a connection one form $A_1 = 0$. We have a second parallel transport map 
${\rm PT}_2$ such that 
for every $\delta \in \Delta$ such that $\delta \neq \nu$ we have 
${\rm PT}_2^\delta = {\rm PT}_1^\delta$. 

Let us use the trivialization over $\nu$ according to which $A_1 = 0$. ${\rm PT}_2^\nu$ described in the same trivialization is characterized by a connection one form $A_2$ determined by a curvature two form $B_2$ which is zero everywhere except for the compact support $s \subset \nu$ shown in figure 1.c. The size of $B_2|_s$ is adjusted in such a way that $\int_D B_2 = 2 \pi$, 
where $D$ is the triangle in ${\rm Sd} (\nu)$ containing $s$. 

All the gluings between bundles over simplices determined by ${\rm PT}_2$ agree with those determined by ${\rm PT}_1$ except for the gluing of the bundles over $\nu$ and 
$\tau$ and $\nu$ and $\nu'$ (the triangle which shares $\tau$ with $\nu$). However, the induced total bundles $\pi_{\Delta, 1}$ and $\pi_{\Delta, 2}$ are inequivalent since their Euler numbers differ by $1$. 

Notice that the parallel transport maps ${\rm PT}_1$ and ${\rm PT}_2$ agree when evaluated on any path contained in the $1$-skeleton of 
${\rm Sd}(\Delta)$. Thus, this discrete set of parallel transport evaluations is not capable of detecting the difference between ${\rm PT}_1$ and ${\rm PT}_2$. It is clear that given any other choice of discrete set of paths, with an appropriate choice of $s$ 
we can fabricate a parallel transport map 
${\rm PT}_2$ which agrees with ${\rm PT}_1$ when restricted to the chosen paths while inducing an inequivalent bundle. 

The gluing between $\pi_\nu$ and $\pi_\tau$ is a change of trivialization of $\pi_\tau$, from the one constructed using ${\rm PT}^\tau$ to the one constructed using 
${\rm PT}^\nu$. As any other change of trivialization it is described by an assignment 
$T_{\nu , \tau}:\tau \to U(1)$. The gluing assignment determined by ${\rm PT}_1$ is 
$T_{\nu , \tau}(x) = {\rm id}$ for all $x \in \tau$, while the gluing determined by 
${\rm PT}_2$ (which obeys compatibility conditions at 
$\partial \tau$) winds around $U(1)$ once. 
For a study of the abelian case see \cite{DZ}.
}

Now we describe (up to scale $\Delta$) 
$[\{ \pi_\nu \to_A \pi_\tau \}_{\tau \subset \partial \nu}
\mbox{(mod. equivalence)}
, A \in {\cal A/G}_{\pi_\nu}]$ in terms of parallel transport maps. 
The macroscopic data needed for such description is 
$\hbox{Data}_\Delta({\rm PT}^\nu)= ( {\rm PT}^\nu_\Delta, W^\nu)$ where
\[
{\rm PT}^\nu_\Delta : {\cal P}_\Delta(\nu) \to G 
\]
is the restriction of ${\rm PT}^\nu$ to 
the discrete semigroup of paths ${\cal P}_\Delta(\nu)\subset {\cal P}(\nu)$ consisting of paths that fit in the $1$-skeleton of ${\rm Sd} (\nu)$. And, 
\[
W^\nu \in \{\mbox{Homotopy class of map } T_{\nu , \tau}: \tau \to G 
\mbox{, with $T|_{\partial \tau}$ fixed} \}_{\tau \subset \nu} , 
\]
where the gluing map $T_{\nu , \tau}$ is constructed form ${\rm PT}^\nu$ following a variation of Barrett's construction \cite{Barrett:1991aj}.  
The result is the following: 
\begin{theorem}[Faithful coarse graining]
\label{EffDescr}
The data $\hbox{Data}_\Delta({\rm PT}^\nu)$ 
characterizes 
\begin{itemize}
\item 
the gluing maps 
$\{ \pi_\tau \to_{{\rm PT}^\nu} \pi_\nu \}_{\tau \subset \partial \nu}$ up to equivalence and 
\item 
the connection modulo gauge $A \in {\cal A/G}_{\pi_\nu})$ up to a microscopical deformation; 
\end{itemize}
where $A, A' \in {\cal A/G}_{\pi_\nu})$ are said to agree up to a microscopical deformation if they can be deformed to the same (singular) semigroup homomorphism by a homotopy of semigroup homomorphisms which preserves $\hbox{Data}_\Delta({\rm PT}^\nu)$. 
\end{theorem}
In the cases relevant for two-dimensioanal and four-dimensional euclidean gravity 
($[ {\rm dim} M = 2, G = U(1)]$ and $[ {\rm dim} M = 4, G = SU(2)]$ respectively) 
$W^\nu$ is characterized by a set of integers, one per codimension one face of $\nu$. 
For many other cases, like the one relevant for three-dimensional euclidean gravity 
($[{\rm dim} M = 3, G = SU(2)]$), $W^\nu$ is trivial. 

This result suggests that if the current spin foam models for 
two-dimensional and four-dimensional gravity 
are modified to include all the relevant macroscopic degrees of freedom 
--the complete collection of macroscopic variables necessary to ensure faithful coarse graining--, 
then they could provide appropriate effective theories at a given scale. 

A spin foam model for two-dimensional gravity which incorporates extra data regarding the bundle structure was given in \cite{Oriti-Rovelli-Speciale}. In the same reference a naive spin foam model which ignored the bundle data was shown to yield unphysical results. An explicit relation between the work presented here and \cite{Oriti-Rovelli-Speciale} will be presented elsewhere.


\section*{Acknowledgements}
This work was partially supported by CONACyT grant 80118.

\end{document}